\def\aa{{A\&A}}

\def\aj{{AJ}}

\def\apj{{ApJ}}

\def\mnras{{MNRAS}}

\documentclass[cup5b]{caps}
\usepackage{mathptmx}
\usepackage{graphicx}
\usepackage{amssymb}
\newcommand{\teff}{$T_{\rm eff}$}
\newcommand{\grad}{g_{\rm rad}}
\newcommand{\zsun}{$Z_\odot$}

\newcommand{\wlr}{{\sc wlr}}
\makeatletter

 \newcommand{\figno}{\scriptsize Fig.}
 \def\fnum@figure{\figno\ \thefigure}

\renewcommand\section{%
  \@startsection {section}{1}{0mm}
    {12.5\p@ \@plus3.5\p@ \@minus2\p@}{0.0011\p@}
    {\Large\normalsc\SFB@hangraggedright}%
}

\renewcommand\subsection{%
  \@startsection{subsection}{2}{0mm}
    {12.5\p@ \@plus3.5\p@ \@minus2\p@}{0.001\p@}
    {\small\normalsc\SFB@hangraggedright}%
}

\def\@startsection#1#2#3#4#5#6{
  \if@noskipsec \leavevmode \fi
  \par
  \@tempskipa #4\relax
  \@afterindentfalse
  \ifdim \@tempskipa <\z@
    \@tempskipa -\@tempskipa \@afterindentfalse
  \fi
  \if@nobreak
    \everypar{}%
    \ifnum#2=\tw@ \vskip 6\p@ \@plus1\p@\fi
    \ifnum#2=\thr@@ \vskip 6\p@ \@plus1\p@\fi
  \else
    \addpenalty\@secpenalty\addvspace\@tempskipa
  \fi
  \@ifstar
    {\@ssect{#2}{#3}{#4}{#5}{#6}}%
    {\@dblarg{\@sect{#1}{#2}{#3}{#4}{#5}{#6}}}}

\def\thefigure {\@arabic\c@figure}

\def\ps@plain{\let\@mkboth\@gobbletwo
  \let\@evenhead\@empty
  \let\@oddhead\@empty
  \def\@evenfoot{\listsize\normalsl\hfill\thepage\hfill}%
  \def\@oddfoot {\listsize\normalsl\hfill\thepage\hfill}%
}

\makeatother

\def\thesection{\arabic{section}.}

\begin{document}

\pagenumbering{arabic}

\author[]{{\large\sc Fabio Bresolin and Rolf-Peter Kudritzki}\\{\rm Institute for Astronomy,
University of Hawaii, Honolulu, HI, USA} \\
\\Invited review at the
Carnegie Observatories Centennial Symposium IV\\ {\rm Origin and Evolution
of the Elements}, Pasadena, 16-21 February 2003}

\chapter{Stellar Winds of Hot Massive Stars \\ Nearby and Beyond the
Local Group}

\begin{center}
\begin{abstract}
\begin{footnotesize}
Photospheric radiation momentum is efficiently transferred by
absorption through metal lines to the gaseous matter in the
atmospheres of massive stars, sustaining strong winds and mass loss
rates. Not only is this critical for the evolution of such stars, it
also provides us with diagnostic UV/optical spectral lines for the
determination of mass loss rates, chemical abundances and absolute
stellar luminosities.  We review the mechanisms which render the wind
parameters sensitive to the chemical composition, through the
statistics of the wind-driving lines. Observational evidence in
support of the radiation driven wind theory is presented, with special
regard to the Wind Momentum-Luminosity Relationship and the blue
supergiant work carried out by our team in galaxies outside of the
Local Group.
\end{footnotesize}
\end{abstract}
\end{center}


\setcounter{section}{1}
\section{\thesection\hspace{3mm} A portrait of mass loss in hot stars}

All stars in the upper H-R diagram, above $\sim10^4\,L_\odot$, are
affected by mass loss via stellar winds (Abbott~1979). Although this
statement includes stars in different evolutionary stages (O and B
main sequence stars, blue and red supergiants, Wolf-Rayet stars,
Luminous Blue Variables), this review will focus on a subset of the
`hot' (\teff\,$\gtrsim$\,8000\,$K$) objects, namely OB main sequence and
blue supergiant stars. In the corresponding regime of luminosity and
effective temperature the radiation field from the stellar photosphere
is strong enough to be able to sustain powerful winds, by transfer of
photon momentum to the metal ions present in the stellar outer
layers. In spite of the low concentration of these ions, when compared
to H and He, the momentum is efficiently distributed to the plasma by
Coulomb interactions, so that the entire stellar envelope is being
accelerated during this process (Springmann \& Pauldrach~1992). It is
intuitive that the chemical composition and overall metallicity of the
stellar outer layers must play a crucial role in determining the
strength of the winds and the rate of mass loss.

The basic description given above is the essence of the radiation
driven wind theory, developed during the course of the last 30 years
in a number of landmark papers. Following the first suggestion by Lucy
\& Solomon (1970), concerning the absorption of radiation in the
ultraviolet resonance lines of ions such as C\,{\sc iv}, Si\,{\sc iv}
and N\,{\sc v}, recently detected at the time as P Cygni profiles from
rocket experiments, the mechanism of wind acceleration by metal line
absorption has been developed further by Castor, Abbott \& Klein
(1975) to include the approximate effect of a larger number of strong
and weak metal lines. The theoretical framework has subsequently been
perfected by the works of Abbott (1982) and Pauldrach, Puls \&
Kudritzki (1986). Approximate analytical solutions for velocity fields
and mass-loss rates within this framework have been developed by
Kudritzki et al.~(1989).  State-of-the-art modeling techniques attain
nowadays a realistic description of radiation driven winds, by solving
the radiation transfer of millions of lines in an expanding
atmosphere, producing synthetic spectra which match the observations
of hot stars from the far-UV to the near-IR to a good degree of
accuracy (Hillier \& Miller 1998; Pauldrach, Hoffmann \& Lennon 2001;
Puls et al. 2003).

\subsection{Scale of the phenomenon}
Based mainly on IR or radio excess and H$\alpha$ emission measurements
quite a broad range in the mass loss rate, $\dot{M}$, has been found
among hot stars, scaling roughly with stellar luminosity as $L^{1.8}$
(de Jager, Nieuwenhuijzen \& van der Hucht~1988; Garmany \&
Conti~1984) during the main sequence and the supergiant phase (it is
actually the modified wind momentum which scales with some power of
the luminosity, as described later). Typical values up to a few times
$10^{-6}\mathcal{M}_\odot$\,yr$^{-1}$ are found among such
stars. Wolf-Rayet stars possess denser winds, with mass loss rates up
to one order of magnitude higher than O stars, while even more extreme
values, up to $10^{-3}\mathcal{M}_\odot$\,yr$^{-1}$, can be found
among Luminous Blue Variables. For central stars of planetary nebulae
$\dot{M}\sim10^{-7}-10^{-9}\mathcal{M}_\odot$\,yr$^{-1}$, in virtue
of their lower luminosities (see the recent review by Kudritzki and
Puls~2000 for further information on stellar wind properties).

\begin{figure}[ht]
 \includegraphics[width=10cm,angle=0]{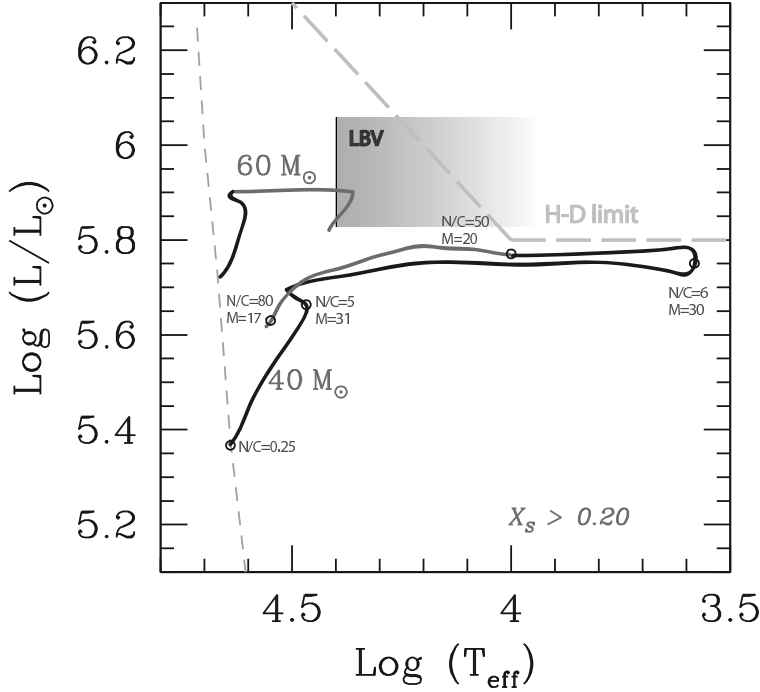}
 \caption{Schematic
 H-R diagram showing the rotating ($V_{\rm i}=300$\,km\,s$^{-1}$) solar
 metallicity 40\,$\mathcal{M}_\odot$ and 60\,$\mathcal{M}_\odot$
 stellar tracks from Meynet \& Maeder (2000). For clarity the tracks
 are plotted up to the point at which the fractional hydrogen mass at
 the surface $X_S=0.2$. The early WN phase is plotted in grey. The
 ZAMS location (short-dashed line) and the Humphreys-Davidson limit
 (long-dashed line) are shown, together with the typical location of
 LBV stars (shaded area). At several positions along the
 40\,$\mathcal{M}_\odot$ track the stellar mass and the surface N/C
 ratio are reported.  } \label{hr} \end{figure}

Evidently such rates of mass loss profoundly influence the evolution
of massive stars (Chiosi \& Maeder 1986). Moreover, winds from massive
stars are relevant for the calculation of chemical yields and their
inclusion in chemical evolution models of galaxies (Maeder 1992;
Portinari, Chiosi \& Bressan 1998). Stellar rotation has also been
shown to significantly affect the evolution of massive stars, in
particular by enhancing mass loss rates and modifying the surface and
wind chemical composition (Meynet \& Maeder 2000). An example of
recent evolutionary models for massive stars is shown in
Fig.~\ref{hr}, where the solar metallicity tracks of Meynet \& Maeder
(2000) for an initial rotational velocity of 300 km\,s$^{-1}$ are
shown for $M=40\,\mathcal{M}_\odot$ and $60\,\mathcal{M}_\odot$. Along
the $40\,\mathcal{M}_\odot$ track the stellar mass and the N/C ratio
are indicated at several stages of evolution.

\setcounter{section}{2}
\section{\thesection\hspace{3mm} Why do stellar winds depend on metallicity?}

Within the galaxies of the Local Group the metallicity varies by more
than one order of magnitude, from metal-poor dwarf irregulars such as
WLM to metal-rich giant spirals such as M31, offering the opportunity
to probe the effects of chemical abundances on the character of the
stellar winds. Expanded possibilities have appeared with the advent of
8-to-10 meter telescopes, which allow us to extend the spectroscopic
study of individual stars even beyond the Local Group.  Quantitative
stellar spectroscopy in external galaxies is becoming important for a
number of projects dealing with stellar abundances, the distance
scale, the evolution of massive stars, the characterization of
supernova progenitors, and the instabilities of massive stars in the
very upper part of the H-R diagram, near the Humphreys-Davidson
limit. The work carried out within our group is mostly concerned with
the first two aspects. For the distance scale work, see the recent
review by Bresolin (2003). Here we will succinctly illustrate how
current observations of hot stars compare with the theoretical
predictions concerning their stellar winds.

\subsection{Theoretical landscape}
As briefly mentioned in the previous section, the winds of hot massive
stars are well described within the framework of the radiation driven
theory. For more details and references on the subject, the interested
reader is referred to the reviews by Kudritzki \& Puls (2000) and
Kudritzki (1998). 

Let us consider a hot and massive star radiating a total luminosity
$L$ out of its photosphere. The outer layers will be accelerated by
radiation pressure through absorption in the metal lines. The maximum
total momentum available from the radiation field is $L/c$, so that we
can expect a dependence of the mechanical momentum flow:
\begin{equation}
 \dot{M}v_\infty=f(L/c) . 
 \end{equation}

As a result of this transfer of momentum from the radiation to the
ions, the wind is accelerated, starting from the photospheric layers,
up to the asymptotic terminal velocity $v_\infty$. The observed
stellar wind velocity fields are customarily parameterized as a
`$\beta$-law', given here in its simplest form:

\begin{equation}
 v(r)=v_\infty \left( 1-\frac{R_\star}{r} \right)^\beta.
 \end{equation}
 
For the hot O stars $\beta\simeq0.8$, therefore the ions are
efficiently accelerated right above the photosphere, while for
later-type A supergiants a more progressive acceleration takes place,
corresponding to $\beta\simeq2-4$.

When considering the hydrodynamics of a stationary line driven wind,
we start from the fundamental equations describing the conservation of
mass

\begin{equation}
 \dot{M}=4\pi r^2 \rho v,
 \end{equation}

\noindent
where $\rho=\rho(r)$ is the local density, together with the equation
of motion

\begin{equation}
 v\frac{dv}{dr} = -\frac{1}{\rho} \frac{dP_{\mbox{\scriptsize gas}}}{dr} - g +\grad^{},
 \label{eom}
 \end{equation}

\noindent
where the radiative acceleration $\grad^{}$ can be expressed as the sum

\begin{equation}
 \grad^{}=\grad^{\mbox{\tiny Th}} + \grad^{\mbox{\tiny  lines}} + \grad^{\mbox{\tiny ff,bf}}.
 \end{equation}

Of these three terms, only the second one (the acceleration due to
line absorption) needs to be considered in more detail, the first one
(Thomson scattering) will be included as a constant reduction of the
gravity $g$, whereas the third one (free-free\,+\,bound-free
transitions) is negligible in the winds of hot stars.

Since the work by Castor et al.~(1975) and Pauldrach et al.~(1986) the
radiative term of the line acceleration is parameterized in terms of a
{\em line force multiplier}, $M(t)$, such that

\begin{equation}
 \grad^{\mbox{\tiny lines}}=C\!F \; \grad^{\mbox{\tiny Th}} \; M(t)
 \end{equation}

\noindent
with $C\!F$ being the correction factor which accounts for the finite
extension of the stellar disk.  The optical depth parameter $t$,
multiplied by the dimensionless line strength $k$ (i.e.~the line
opacity divided by the Thomson opacity $n_e \sigma_e$, see Kudritzki
and Puls~2000), gives the optical depth in a given line transition in
a supersonically expanding atmosphere:

\begin{equation}
 \tau=k\,t
 \label{tau}
 \end{equation}
\begin{equation}
 t=n_e \sigma_e \frac{v_{\rm therm}}{dv/dr}.
 \label{t}
 \end{equation}

In its simplest form (see Abbott~1982, Pauldrach et al.~1986 and
Kudritzki~2002 for a more accurate description) the line force
multiplier is then expressed as

\begin{equation}
 M(t) \propto N_{\rm eff} \, t^{-\alpha} , 
 \label{gradeq}
 \end{equation}

\noindent
thus depending on the effective number of metal lines driving the
wind, $N_{\rm eff}$, and on the optical depth through the line force
parameter $\alpha$. Solving the equation of motion (\ref{eom}) with
this parameterization of the line force one obtains scaling relations
for the mass-loss rate and the terminal velocity (Kudritzki et
al.~1989):

\begin{equation}
 \dot{M} \sim (N_{\rm eff} \, L)^{1/\alpha} \, \big[M_\star(1-\Gamma)\big]^{1-1/\alpha}
 \hspace{7mm} v_\infty \sim \frac{\alpha}{1-\alpha} \, v_{\rm esc}.
 \label{scale}
 \end{equation}

Note that $M_\star(1-\Gamma)$ is the stellar {\em effective mass} (which
accounts for the ratio $\Gamma$ of Thomson scattering to gravitational
acceleration). 

\begin{figure}[ht]
 \includegraphics[scale=0.75]{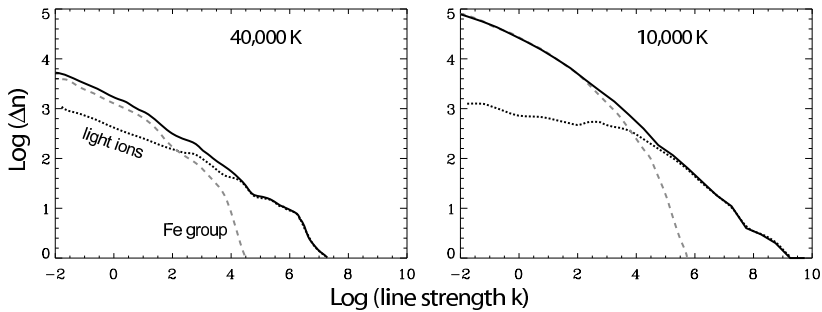}
 \caption{Line distribution functions for \teff\,=\,40,000$\,K$ (left)
 and \teff\,=\,10,000$\,K$ (right) and solar composition, plotted
 separately for iron group elements (dashed lines) and lighter ions
 (dotted lines). The full line represents the total distribution
 (adapted from Puls et al. 2000). } 
 \label{df}
 \end{figure}

The origin of the $M(t)$ expression used in the proportionality
(\ref{gradeq}) lies in the statistics of the strengths of the
wind-driving lines, which can be described by a simple {\em line
strength distribution function}. It is found that, to a good
approximation, the number of lines of a given strength $k$ follows a
power-law (Fig.~\ref{df}):

\begin{equation}
 n(k)d(k) \propto k^{\alpha-2}dk
 \hspace{2cm} (0<\alpha<1;
 \hspace{5mm} 1<k<k_{\rm max}).
 \end{equation}

The exponent $\alpha$, which determines the slope of the distribution
function, is mostly set by the laws of atomic physics (the
distribution function of oscillator strengths), and is found to vary
between 0.5 and 0.7 (for the hydrogen Lyman-series one would obtain
$\alpha$\,=\,2/3). $N_{\rm eff}$ (see equations \ref{gradeq},
\ref{scale}) is related to the normalization of the line strength
distribution function. Also note that a power-law is only an
approximation and that there is a slight curvature in the distribution
function.

The line distribution function plays a primary role in the mechanisms
governing radiation driven winds (see Puls, Springmann \& Lennon 2000
for an in-depth study), and it is worthwile spending a few additional
words in relation to the type of ions which are most effective in
driving the wind. It is clear that such ions will differ among stars
of different chemical composition and/or effective temperatures, since
the predominant ionizations stages will change accordingly.

What kind of lines are driving the wind?  The most prominent metal
line transitions are located in the far-UV and UV spectral ranges. As
the emitted stellar flux peaks at smaller frequencies with decreasing
\teff, so does the spectral range containing the lines contributing
the most at driving the wind. The calculations by Abbott (1982) and
Puls et al.~(2000) show that lines of high-ionization stages of O, N,
P and other heavy elements, located approximately between 800 and
1200\,\AA, are the dominant source of acceleration in the hottest
(40,000$\,K$) stars. At lower temperatures (10,000--20,000$\,K$) lower
ionization species (such as Fe\,{\sc ii}, Fe\,{\sc iii}, Mg\,{\sc ii},
Ca\,{\sc ii}), having transitions at longer wavelengths, become
predominant.  The relative importance of CNO, Fe group and other
elements changes with \teff\/ as well. For solar composition C, N and
O dominate over iron group elements for \teff$\;>\;$\,25,000$\,K$, and
down to lower temperatures as the overall metallicity decreases (Vink,
de Koter \& Lamers 2001).  As the right panel in Fig.~\ref{df} shows,
at lower temperature the number of iron group lines present, mostly
Fe\,{\sc ii}-{\bf {\sc iii}}, increases, determining a steepening of
the line distribution function, and a consequent decrease in $\alpha$.

We can now answer our question regarding the origin of the metallicity
dependence of the mass loss, based on the understanding of the line
statistics briefly discussed so far. There are two main effects:

$(i)$ to first order, the line strengths of the individual metal lines
driving the wind are proportional to metallicity

\begin{equation}
 k \sim \frac{Z}{Z_\odot}\,k_\odot
 \end{equation}

\noindent
so that the distribution functions in Fig.~\ref{df} shift
horizontally in the plotted \mbox{log--log} plane. Consequently, the 
normalization over the range $1<k<k_{\rm max}$ changes with metallicity,
affecting the effective number of lines driving the wind:

\begin{equation}
 N_{\rm eff} \propto Z^{1-\alpha}.
 \label{neff}
 \end{equation}

$(ii)$ a metallicity dependence of $\alpha$, coming from the curvature
of the line distribution function, which tends to become steeper at
high line strengths (see Puls et al.~2000 for the details). It is
important to note that the local acceleration in the wind comes mostly
from lines with $\tau \simeq 1$. This means that strong winds are driven
by lines with smaller line strengths, whereas weak winds rely on the
acceleration of the fewer lines with large line strengths.

To conclude, the predicted scaling relations for the metallicity
dependence of the wind parameters, valid approximately in the
range $0.1 < Z/Z_\odot < 3$, are:

\begin{equation}
 \dot{M} \sim Z^{(1-\alpha)/\alpha}=Z^m
 \end{equation}

\noindent
with $m=0.5-0.8$ for O- and B-type stars, and

\begin{equation}
 v_\infty \sim Z^{0.13-0.15}
 \end{equation}

\noindent
(Kudritzki, Pauldrach \& Puls 1987; Leitherer, Robert \& Drissen
1992). While the effects on the mass-loss are mostly a direct result
of $(i)$ through Eqs.~(\ref{scale}), (\ref{neff}), the metallicity
dependence of $v_\infty$ is solely caused by $(ii)$ and
Eq.~(\ref{scale}). At very low metallicities (below
$10^{-2}\,Z_\odot$) these relations break down, and the depth
dependence of the line force multipliers must be taken into account
(Kudritzki 2002). We also conclude from Fig.~\ref{df} that a change of
the relative abundance ratio of $\alpha$ elements to Fe group elements
will have an additional effect.

\setcounter{section}{3}
\section{\thesection\hspace{3mm} Does Nature conform to theory?}\label{wlrsect}

The first, obvious observational test of the effects of metallicity on
the stellar winds of massive stars concerns the terminal
velocities. These are easily measured from the P Cygni profiles of
resonance lines in the UV, e.g.~C\,{\sc iv}\,$\lambda1550$ and
Si\,{\sc iv}\,$\lambda1400$. A clear decrease of $v_\infty$ is found
when comparing the terminal velocities measured in the Milky Way with
those in the Magellanic Clouds, especially in the lower metallicity
SMC (Garmany \& Conti 1985; Kudritzki \& Puls 2000). Direct
comparisons of O-star P Cygni profiles at high (Galactic) and low
(SMC) metallicity can be found, for example, in Leitherer et
al.~(2001) and Heap, Hubeny \& Lanz (2001).

Concerning the mass loss rates, their derivation is more model
dependent. The bulk of the measurements available for massive stars
comes from H$\alpha$ line profile fits. However, when correlating
$\dot{M}$ with, for example, the luminosity $L$, a significant scatter
is found, as a consequence of the dependence on the effective mass
(see Eq.~\ref{scale}). This difficulty vanishes, at least
theoretically, when considering the wind momentum through
Eq.~(\ref{scale}), since $v_{\rm esc}$ depends on the square root of the
effective gravitational potential:

\begin{equation}
 \dot{M}v_\infty \propto \frac{(N_{\rm
eff}\,L)^{1/\alpha}}{R_\star^{0.5}} \, \big[M_\star(1-\Gamma)\big]^{3/2-1/\alpha}.
 \label{this}
 \end{equation}

The fortuitous nulling of the exponent for the effective mass
(since $\alpha\simeq 2/3$ for hot stars) then relates the {\em modified
wind momentum}

\begin{equation}
 D_{\rm mom} = \dot{M} v_\infty (R_\star/R_\odot)^{0.5}
 \end{equation}

\noindent
to the stellar luminosity in a simple Wind Momentum-Luminosity
Relationship ({\sc wlr}):

\begin{equation}
 \log D_{\rm mom} = a + b\, \log\frac{L}{L_\odot}
 \label{wlr_eq}
 \end{equation}

\noindent
(see Kudritzki 1988 or Kudritzki \& Przybilla 2003 for a recent
detailed derivation), which constitutes the basis of our program to
use the winds of massive stars in external galaxies to determine their
distances (an independent method valid for blue supergiants, the
Flux-weighted Gravity--Luminosity Relationship ({\sc fglr}) has been
recently proposed by Kudritzki, Bresolin \& Przybilla 2003).

\begin{figure}[ht]
 \includegraphics[scale=0.6]{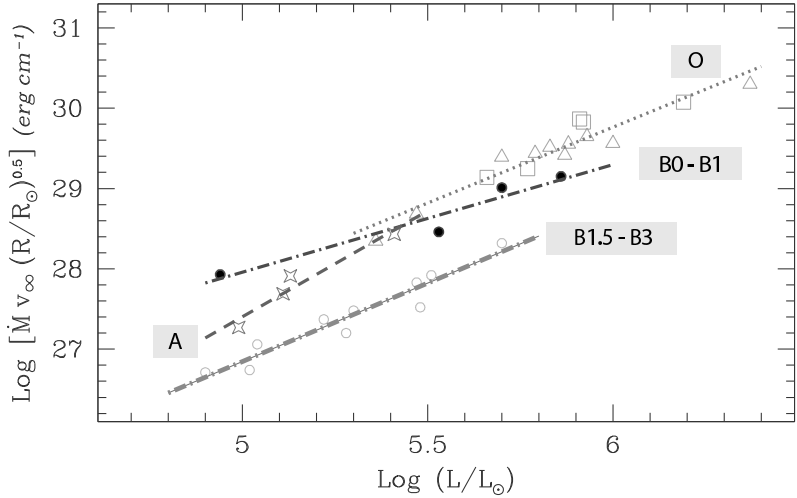} \caption{Spectral type
 dependence of the \wlr\/ for Galactic supergiant stars. Different
 symbols are used for different spectral type ranges, and the
 corresponding linear regression lines are drawn (from
 Kudritzki et al.~1999).}  \label{wlr_fig} \end{figure}

From equation (\ref{this}) it is evident that $a$ and $b=1/\alpha$ in
equation (\ref{wlr_eq}), which are to be derived from observations for
an empirical {\sc wlr}, are a function of the predominant ionization
stages in the stellar atmosphere (spectral type) and of the
metallicity. The slope and zero-point of the \wlr\/ are in fact found
to differ for O, B and A supergiant stars by Kudritzki et al.~(1999),
as illustrated in Fig.~\ref{wlr_fig}.

Combining the metallicity effect on $\dot{M}$ and $v_\infty$ discussed
in the previous section, the theoretical dependence of the modified
wind momentum on metallicity becomes

\begin{equation}
 D_{\rm mom} \sim Z^{0.6-0.8} .
 \label{that}
 \end{equation}

Puls et al.~(1996) and Herrero, Puls \& Najarro (2002) have shown that
the Galactic O stars follow the \wlr\/ predicted by the radiation
driven theory rather convincingly. Data on B and A supergiants, as
well as on lower metallicity massive stars which would allow a test of
equation (\ref{that}), remain scant. Puls et al.~(1996) indeed found a
decrease in the wind momentum for a small sample of O stars in the
Magellanic Clouds compared to similar stars in the Milky Way, in rough
agreement with the theoretical expectations. Vink et al.~(2001) have
also found reasonable agreement between predicted and observed mass
loss rates of O-type stars in the LMC and SMC, if the adopted
metallicity is 0.8\,\zsun\/ and 0.1\,\zsun, respectively.

\setcounter{section}{4}
\section{\thesection\hspace{3mm} Looking beyond the Local Group}
Among the most exciting developments in stellar research in recent
years has been the possibility of quantitative studies of individual
stars in external galaxies located even beyond the Local Group, thanks
to observations with 8\,m-class telescopes. When working at
$V\simeq19-20$ or fainter in a galaxy several Mpc away, the natural
targets become the visually brightest mid-B to early-A supergiants and
hypergiants (with the occasional Luminous Blue Variable). Reaching
absolute magnitudes \mbox{$M_V\simeq-8$} to $-9$, they can be studied
spectroscopically out to $10-15$ Mpc. Such stars hold the promise of
delivering important information on stellar abundance patterns, a
welcome complement to H\,{\sc ii} region studies, and extragalactic
distances (via the \wlr\/ and the {\sc fglr}) even at a moderate
$R\simeq1000-2000$ spectral resolution.

\begin{figure}[hb]
 \includegraphics[scale=0.65]{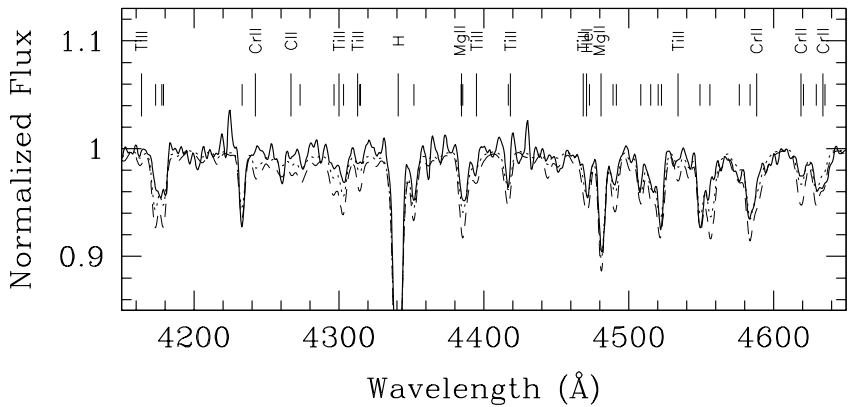} \caption{Portion of the
 rectified blue VLT/FORS spectrum of one A0\,Ia star in
 NGC~300. Principal spectral lines are identified at the
 top. Synthetic models are shown for $Z=Z_\odot$ (dashed line) and
 $Z=0.5\,Z_\odot$ (dotted line).}  \label{d13} \end{figure}

Pioneering results concerning quantitative stellar spectroscopy at
such distances have been obtained recently within our collaboration,
taking advantage of the multi-spectrum capabilities offered at the
Very Large Telescope with the FORS instrument. Supergiant stars in
NGC~3621, a late-type spiral with a Cepheid distance of 6.7\,Mpc,
studied by Bresolin et al.~(2001), remain the farthest {\em normal}
stars for which information regarding the stellar wind and abundances
have been obtained so far. The NLTE spectral synthesis of two A-type
supergiants in NGC~3621 reveals a sub-solar overall metallicity
(Bresolin et al.~2001; Przybilla 2002), and indicates that the
internal accuracy in abundance, even at the moderate FORS resolution,
is $\sim0.2$ dex.

Additional blue supergiants have been investigated spectroscopically
in NGC~300, at a distance of 2\,Mpc, by Bresolin et al.~(2002a). This
work is part of a larger project aiming at improving the accuracy of
stellar candles used to measure distances of nearby galaxies. The
study of the blue supergiants, in particular, will provide stellar
chemical compositions, needed to test the Cepheid P--L relation at
varying metallicities, and at the same time independent distances to
the parent galaxies via the \wlr\/ and the {\sc fglr}, once calibrated
in nearby galaxies (Bresolin 2003).

\begin{figure}[ht]
 \includegraphics[scale=0.72]{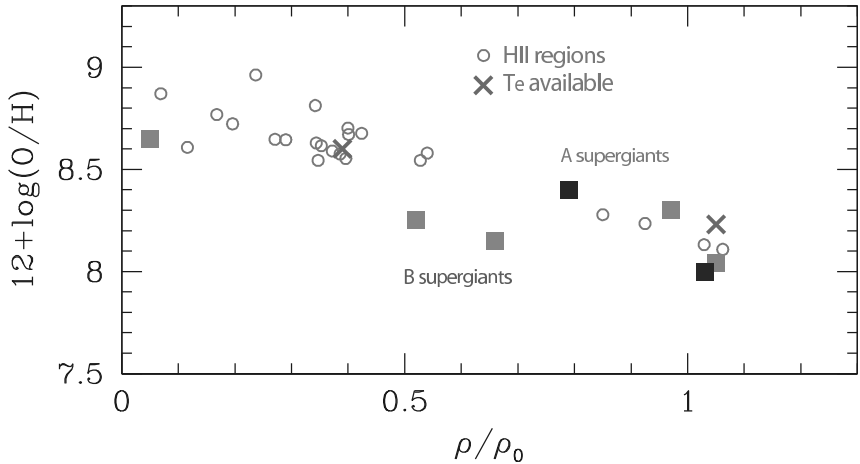} \caption{Preliminary
 comparison of the nebular (circles) and stellar (squares) abundance
 gradients derived for NGC~300, expressed in terms of the fractional
 isophotal radius. The H\,{\sc ii} region abundances have been
 estimated from the semi-empirical $R_{23}$\,=\,([O\,{\sc ii}] +
 [O\,{\sc iii}])/H$\beta$ calibration by Kobulnicky, Kennicutt \&
 Pizagno (1999), except for two regions (crosses), for which a direct
 abundance determination has been made possible. } \label{grad}
 \end{figure}

The abundance diagnostic power of the A supergiant VLT spectra in
NGC~300 is illustrated in Fig.~\ref{d13}, where $Z/Z_\odot\simeq0.5$
is derived for an A0\,Ia star from a comparison with synthetic
spectra. A similar work has been carried out for early B-type
supergiants from the same set of spectra by Urbaneja et
al.~(2003). This allows us a comparison between stellar and nebular
abundances in NGC~300, using published results on the O/H abundance of
H\,{\sc ii} regions (Deharveng et al.~1988), and adopting [O/H]=[M/H]
for A stars, where M/H is the mean stellar metallicity
(Fig.~\ref{grad}).

Only for very few external spiral galaxies such kind of comparison has
insofar been carried out, namely the Local Group members M33
(Monteverde et al.~1997, 2000) and M31 (Venn et al.~2001; Smartt et
al.~2001; Trundle et al.~2002; see also K.~Venn's contribution on
dwarf irregular galaxies at this conference). However, the importance
of such comparisons cannot be overstated, because of the need for a
check on the nebular abundances, which in the case of most spiral
galaxies relies on empirical {\em strong line} methods, which can
provide systematic errors on the estimated abundances amounting to
factors of two or three (Kennicutt, Bresolin \& Garnett~2003). While
the result for NGC~300 is still very preliminary, and based on a small
number of stars, it suggests a rough agreement between stars and
H\,{\sc ii} regions, even though the O/H scale for the latter is
dependent upon which empirical calibration one chooses to adopt.

\subsection{Stellar winds in supergiants outside the Local Group}
High-resolution studies of individual blue supergiants outside of the
Milky Way and the Magellanic Clouds have been made in M33 (McCarthy et
al.~1995), M31 (McCarthy et al.~1997, Venn et al.~2000), NGC~6822
(Venn et al.~2001) and WLM (Venn et al.~2003). The winds of some of
these stars have also been analyzed, and their strength roughly
corresponds with the theoretical expectations, based on the measured
luminosities and abundances. However, a systematic investigation of
the wind properties of a sizeable sample of blue supergiants in a
single galaxy has yet to come. This is necessary for understanding the
parameters affecting the strength of stellar winds, in particular
their dependence on metallicity. The NGC~300 supergiant sample
presented by Bresolin et al.~(2002a), soon to be complemented by
objects observed at the VLT in another Sculptor member, NGC~7793, give
us the chance to partly remedy the situation, although at such
distances lower spectral resolutions must be used.

Six A supergiants in NGC~300 have been analyzed with the unblanketed
version of the {\sc fastwind} code described by Santolaya-Rey, Puls \&
Herrero (1997) to measure mass loss rates from the H$\alpha$ line
profiles and gravities from H$\gamma$. $\dot{M}$ was found to span the
range $1.4\times10^{-8} -
2.3\times10^{-6}\mathcal{M}_\odot$\,yr$^{-1}$, with luminosities
$\log (L/L_\odot)=4.8 - 5.7$ ($M_V=-6.8$ to $-8.6$).

\begin{figure}[ht]
 \includegraphics[scale=0.77]{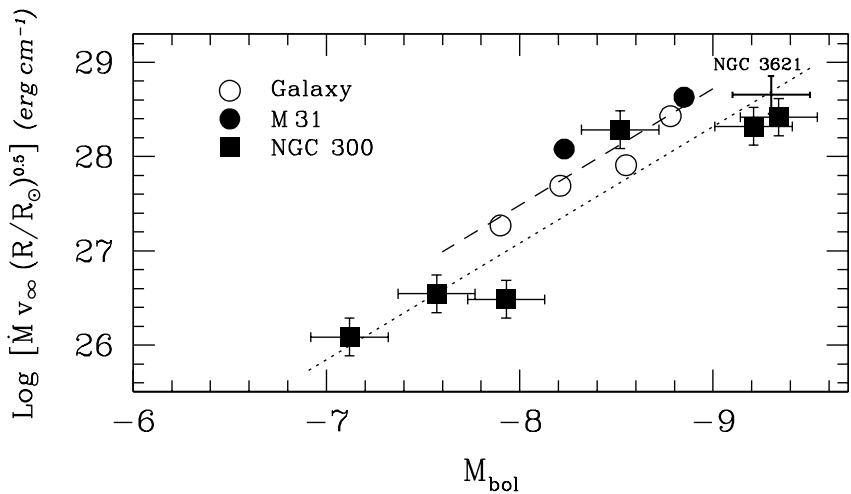}
 \caption{The A-type supergiant \wlr, including objects from the Milky
Way, M31, NGC~300 and NGC~3621. The linear fit to the Galactic and M31
stars is shown by the dashed line. A theoretical scaling factor is
applied to provide the expected relation at $Z/Z_\odot$\,=\,0.4 (dotted
line), the mean metallicity found for the NGC~300 and NGC~3621 stars
included in the plot.}  
 \label{n300wlr} 
 \end{figure}

The resulting \wlr\/ is displayed in Fig.~\ref{n300wlr}, where results
for A supergiants in the Milky Way, M31 (Kudritzki et al.~1999) and
NGC~3621 (Bresolin et al.~2001) are included. As can be seen, the
$D_{\rm mom}-Z$ scaling relation (Section~\ref{wlrsect}) applied to
the Galactic regression line for the mean metallicity of the NGC~300
stars (\mbox{$Z\simeq 0.4\,Z_\odot$}, dotted line) provides a
reasonable fit to the data. Some discrepancies are present, but
our general conclusion is that theory and observations are in fair
agreement. The analysis of a larger number of stars, including the
effects of blanketing, is needed.

\setcounter{section}{5}
\section{\thesection\hspace{3mm} Multi-wavelength studies}
The importance of the multi-wavelength approach in the study of
massive star winds and abundances has been clearly illustrated by
Taresch et al.~(1997), who analyzed far-UV (ORFEUS), UV (IUE) and
optical spectra of the Galactic star HD\,93129A (recently included in
the newly defined O2\,If$^*$ spectral class by Walborn et
al.~2002). While the UV spectral range has in general been much more
accessible than the far-UV, the amount of information regarding
temperatures and abundances which can be collected from the UV alone
is quite limited. Therefore, the iron abundance was estimated from
spectrum synthesis of the plethora of iron lines present in the UV to
be roughly solar (see Haser et al.~1998 for a similar approach applied
to Magellanic Cloud O stars).  This result was complemented by the
measurement of optical lines to derive CNO abundances (finding a
2\,$\times$\,solar overabundance for N, and a $\sim 5\,\times$
depletion for C and O), and by the access to additional ionization
stages of several elements (e.g.~unsaturated lines of C\,{\sc iii},
N\,{\sc iii-iv}, O\,{\sc vi}, S\,{\sc vi}, P\,{\sc v}), which become
important in contraining the abundances and stellar effective
temperatures, as well as the mass loss rates.

The effectiveness of this technique has been recently demonstrated by
Crowther et al. (2002a), who analyzed FUSE far-UV wind-affected metal
lines of four Magellanic Cloud O supergiants, together with
IUE\,+\,HST UV and optical data.  The far-UV spectra have been crucial
in fixing the \teff's, which resulted systematically 5-7,000\,$K$
lower than determined previously from unblanketed, plane-parallel
models (see also Puls et al.~2003; Bianchi \& Garcia 2002). Mass loss
rates are also substantially revised downwards, as a consequence of
lower luminosites.  As in the case of HD\,93129A studied by Taresch et
al., strong N enrichment has been detected, together with modest C
depletion, adding another piece of evidence for mixing processes
affecting the surface chemical composition of evolved massive stars
(see also Lennon, Dufton \& Crowley~2003).

As a final remark, it is worth emphasizing the importance of the
modeling techniques used to derive the stellar parameters, currently
relying on the full line blanketing and spherical extension treatment
in non-LTE of the radiative transfer equation in the expanding
atmospheres. Unified model atmosphere codes, such as {\sc cmfgen}
(Hillier \& Miller 1998) and {\sc wm}-basic (Pauldrach et al.~2001),
are now routinely employed to investigate the stellar properties of
massive stars.

\setcounter{section}{6}
\section{\thesection\hspace{3mm} Additional matters}

\subsection{Wolf-Rayet stars}
Wolf-Rayet stars are spectacular manifestations of the profound
effects of mass loss on the evolution of massive stars. The removal of
the outer layers of the stars progressively reveals, towards the late
evolutionary stages, the products of H-burning (WN stars) and,
subsequently, He-burning (WC stars). Because of the mechanism
responsible for this effect, i.e.~the transfer of photon momentum to
the gas via metal line absorption, the efficiency of this peeling
process is highly dependent on metallicity. First of all, the minimum
progenitor mass required for W-R star formation decreases with
increasing metallicity (Maeder \& Meynet 1994), ranging
observationally from 20--25\,$\mathcal{M}_\odot$ in the Milky Way to
70\,$\mathcal{M}_\odot$ in the SMC (Massey et al.~2000,
2001). Moreover, the WC/WN number ratio, which measures the efficiency
of the evaporation process, increases from virtually zero at SMC
abundance to almost one at super-solar abundance, as in the case of
M31 (Massey~2003).

The high mass loss of Wolf-Rayet stars is manifested by the strong H,
He and metal emission lines present in their spectra. This facilitates
their detection with narrow-band {\em on-off} imaging techniques, and
provides the spectroscopist with a means to probe detailed chemical
abundance patterns in their outer layers, which invariably reflect
their advanced evolutionary status. In this regard, besides
quantitative work in the Milky Way, the Magellanic Clouds and a few
Local Group galaxies (e.g.~Herald, Hillier \& Schulte-Ladbeck 2001;
Smartt et al.~2001; Crowther et al.~2002b), objects at much larger
distances have started to be analyzed. Bresolin et al.~(2002b) have
presented chemical abundance patterns in a WN11 star they discovered
in NGC~300. In this galaxy Schild et al.~(2003) list nearly 60 W-R
stars, based on a VLT imaging survey, and analyze the spectra of two
WC stars. According to these authors there are at least a dozen W-R
stars or W-R star candidates brighter than $V\simeq19$ (the magnitude
of the WN11 studied by Bresolin et al.~2002b) in this galaxy alone,
opening up the possibility for quantitative studies of a large number
of emission-line stars outside the Local Group.

\subsection{Starbursts and the high-redshift universe}
It is of consolation to part of the hot-star community that the same
kind of analyses and modeling used for nearby single stars can be
successfully used for the understanding of stellar populations and
chemical abundances in distant galaxies. Such is the case in the study
of starburst events at distances where only the integrated light from
a composite stellar population can be measured, where strong UV wind
and photospheric lines can help constrain the properties of the
emitting regions. Population synthesis models have become popular in
order to disentangle several characterizing properties of the
star-forming regions, such as the initial mass function, the star
formation history, the age and the chemical abundance. Recent examples
of the application of this technique are given by Gonzalez Delgado et
al.~(2002) for the metal-rich starburst galaxy NGC~3049, and Leitherer
et al.~(2001) for NGC~5253. In the latter case the main UV spectral
features are reasonably well reproduced by synthetic models based on
UV stellar libraries constructed from metal-poor Magellanic Cloud O
stars, rather than Galactic ones. A similar result has been obtained
by the same authors, as well as by Heap et al.~(2001), for the
well-known lensed galaxy MS1512-cB58. However, we have already run
short of UV stellar templates, as we currently have no other
metallicities represented in the synthetic models other than those of
the Galactic and the Magellanic Cloud stars. Besides, B stars are
included only for the Galactic case. The solution to these
difficulties is represented by the use, instead of observed stellar
spectra, of model UV spectra, where the metallicity can be varied at
will. The reliability of the stellar models must of course be first
tested against the existing templates. The outcome in the near future
will be a better knowledge of the properties, in particular of the
chemical abundances, of high-redshift star forming regions and
primordial galaxies.


\end{document}